\documentclass[conference]{IEEEtran}
\IEEEoverridecommandlockouts
\usepackage{cite}
\usepackage{amsmath,amssymb,amsfonts}
\usepackage{algorithmic}
\usepackage{graphicx}
\usepackage{textcomp}
\PassOptionsToPackage{usenames,dvipsnames,svgnames}{xcolor}  
\usepackage{tikz}
\usetikzlibrary{arrows,positioning,automata}
\usetikzlibrary{shapes.symbols,patterns}
\usetikzlibrary{shapes.geometric,fit}
\usetikzlibrary{calc}
\usetikzlibrary{decorations.pathreplacing,decorations.markings,shapes.geometric}
\tikzset{naming/.style={align=center,font=\small}}
\tikzset{antenna/.style={insert path={-- coordinate (ant#1) ++(0,0.25) -- +(135:0.25) + (0,0) -- +(45:0.25)}}}
\tikzset{station/.style={naming,draw,shape=dart,shape border rotate=90, minimum width=10mm, minimum height=10mm,outer sep=0pt,inner sep=3pt}}
\tikzset{mobile/.style={naming,draw,shape=rectangle,minimum width=12mm,minimum height=6mm, outer sep=0pt,inner sep=3pt}}

\def\BibTeX{{\rm B\kern-.05em{\sc i\kern-.025em b}\kern-.08em
    T\kern-.1667em\lower.7ex\hbox{E}\kern-.125emX}}
\begin{document}

\title{Energy Balancing for Robotic Aided Clustered Wireless Sensor Networks Using Mobility Diversity Algorithms\\
\thanks{}
}

\author{\IEEEauthorblockN{Daniel Bonilla Licea$^{*}$, Edmond Nurellari$^{\dagger}$, and Mounir Ghogho$^{*,\ddagger}$}
\IEEEauthorblockA{$*${International University of Rabat, FIL, TICLab, Morocco} \\
{$\dagger$ School of Engineering, University of Lincoln, UK }\\
{$\ddagger$ School of Electronic and Electrical Engineering, University of Leeds, UK }\\
daniel.bonilla-licea@uir.ac.ma, enurellari@lincoln.ac.uk, m.ghogho@ieee.org}
}

\maketitle

\begin{abstract}
We consider the problem of energy balancing in a clustered wireless sensor network (WSN) deployed randomly in a large field and aided by a mobile robot (MR). The sensor nodes (SNs) are tasked with monitoring a region of interest (ROI) and reporting their test statistics to the cluster heads (CHs), which they subsequently report to the fusion center (FC) over a wireless fading channel. To maximize the lifetime of the WSN, the MR is deployed to act as an adaptive relay between a subset of the CHs and the FC. To achieve this we develop a $multiple-link$ mobility diversity algorithm (MDA) executed by the MR that will allow to compensate simultaneously for the small-scale fading at the established wireless links (i.e., the MR-to-FC as well as various CH-to-MR communication links). Simulation results show that the proposed MR aided technique is able to significantly reduce the transmission power required and thus extend the operational lifetime of the WSN. We also show how the effect of small-scale fading at various wireless links is mitigated by using the proposed $multiple-link$ MDA executed by a MR equipped with a single antenna.
\end{abstract}


\section{Introduction}
\label{sec:intro}
Monitoring a region of interest (ROI) is one of the most important applications of wireless sensor networks (WSNs) \cite{1}, \cite{2}. Multiple low-cost sensor nodes (SNs) are often spatially deployed over a large ROI to observe different events and estimate parameters of interest. In general, the SNs process the local observations and report back to a fusion center (FC) that optimally combines the individual reports to reach a global decision. Being geographically dispersed to cover large areas, the SNs are  constrained in both bandwidth and power. To allow a $low-latency$ WSN  when the ROI is very large, the WSN is divided into multiple clusters to manage the large number of SNs needed to provide reliable coverage (e.g., see Fig. \ref{Figure555}). Each cluster head (CH) receives data from each SN within the cluster, which subsequently reports to the FC where the ultimate decision is taken. 

The framework of centralized decision for a single FC (i.e., single CH) network configuration has been extensively studied in \cite{3,4,5,44}, to name but a few references. There are some recent publications \cite{6,7} (in the context of estimation) and \cite{8} (in the context of detection) that considered the effect of $inter\hspace{-0.03cm}-\hspace{-0.03cm}sensor$ collaboration on the WSN performance; after the collaboration stage, the SNs (which in general can be a subset of all SNs) report to a FC where the final decision is made. While the authors in \cite{6} claim to reduce the FC control overhead, \cite{7} derives the optimum power allocation scheme for a given maximum total network power budget in order to improve the estimation quality. 

Now, clustered WSNs \cite{9} has been extensively studied in various contexts such as energy management \cite{10, 11} and fusion rules design\cite{12}. In the context of clustering algorithms, the authors in \cite{13} propose a $d-hop$ cluster partitioning to deal with the load imbalance among CHs.  In this paper, we adopt the network configuration in \cite{5}, i.e., we consider equal-sized clusters of SNs (e.g., \cite{14}) that report their information on a regular basis to a FC. For the CHs that are too far from the FC and when communication link between the CH and the FC is poor, a mobile robot (MR) is deployed to act as an adaptive relay. We propose a $multiple-link$ MDA to extend the operational lifetime of the WSN and to deal with the imbalanced load among the CHs. During the execution of the  $multiple-link$ MDA the MR will move small distances (on the order of one wavelength). The transmission (CHs-to-MR as well as MR-to-FC) links are assumed to experience shadowing and multipath fading. We show that the proposed MDA effectively deals with the energy imbalance in a cluster WSN.
\section{System Model}
\label{system model}
Consider the problem of monitoring a large ROI by a WSN consisting of a FC, $M$ spatially distributed SNs that are networked in $N$ equal-sized clusters and a MR; all equipped with a single antenna. The wireless channels CHs-to-MR and the channel MR-to-FC are assumed to experience shadowing as well as small scale fading\footnote{The small scale fading is assumed to be time-variant with a coherence time $\tau$.}.  Since most of the WSNs are bandwidth constrained, we assume narrow band communications so that the communication channels can be modeled as  non-frequency selective. The case where the spatially distributed CHs report to the FC  via a dedicated parallel access channel (PAC) is investigated in e.g., \cite{11, 12}. Here, we propose to deploy a single MR to act as an adaptive relay for forwarding the test statistics from the CHs to the FC. So, at the MR (positioned at point $\mathbf{p}(t)$), the test statistic received from the $j$th CH at time $t$ is:
\begin{equation}
\label{eq:2.1}
\hat T_j(t)=\left(\frac{s(\mathbf{p}(t),\mathbf{q}_j)h(\mathbf{p}(t),\mathbf{q}_j,t)}{\|\mathbf{p}(t)-\mathbf{q}_j\|_2^{\alpha/2}}\right)T_j(t)+n(t)
\end{equation}
where $\mathbf{q}_j$ is the position of the $j$th CH, $s(\mathbf{p}(t),\mathbf{q}_j)$ represents the shadowing which is modeled by a lognormal random variable whose normalized spatial correlation function is exponential, $T_j(t)$ is the test statistic transmitted from the $j$th CH, and $n(t)\sim \mathcal{N}(0,{\sigma_{i}}^{2}))$; $h(\mathbf{p},\mathbf{q}_j,t)$ represents the small scale fading assumed to follow Jakes' model, i.e., its normalized spatial correlation is:
\begin{equation}
\label{eq:2.2}
\rho(\mathbf{p},\mathbf{q})=\mathbb{E}\left[h(\mathbf{o},\mathbf{p},t)h^*(\mathbf{o},\mathbf{q},t)\right]=J_0\left(\frac{2\pi\|\mathbf{p}-\mathbf{q}\|}{\lambda}\right)
\end{equation}
where $\mathbf{o},\mathbf{p},\mathbf{q}\in\mathbb{R}^2$ are arbitrary points in the space; $J_0(\cdot)$ is the Bessel function of first order and zeroth degree while $\lambda$ is the wavelength of the carrier used for the transmission. In addition, the small-scale fading is assumed to remain constant over the coherence time $\tau$ (i.e.,  $h(\mathbf{p},\mathbf{q}_j,t)$= $h_k(\mathbf{p},\mathbf{q}_j)$ and for $t\in[k\tau,(k+1)\tau)$), $h_{k_1}(\mathbf{p},\mathbf{q}_j)$ and $h_{k_2}(\mathbf{p},\mathbf{q}_j)$ are assumed statistically independent for $k_1\neq k_2$. 

Now, without loss of generality, we assume that the distance between the CHs is significantly larger compared to $\lambda$ and so, for $j\neq i$, $h_k(\mathbf{p},\mathbf{q}_j)$ and $h_k(\mathbf{p},\mathbf{q}_i)$ are considered to be statistically independent. 

For notational convenience we denote the position of the FC as $\mathbf{q}_0$.

Finally, to satisfy a certain average reference power $P_{ref}$ at the receiver, the CHs and the MR use transmit power control mechanism. At the $j$th CH, the average transmit power is:
\begin{equation}
\label{eq:2.3}
P_j=\frac{\|\mathbf{p}(t)-\mathbf{q}_j\|_2^{\alpha}P_{ref}}{s_j^2(\mathbf{p}(t),\mathbf{q}_j)\left|h_k(\mathbf{p}(t),\mathbf{q}_j)\right|^2}
\end{equation}
where $t\in[k\tau,(k+1)\tau)$, and $\alpha$ is the path loss coefficient.

\begin{figure}[!ht]
\begin{center}
\end{center}\vspace{-0.25cm}
\centerline{{\includegraphics[clip, trim ={0mm 80mm 10mm 10mm},width=80mm ,height=62mm]{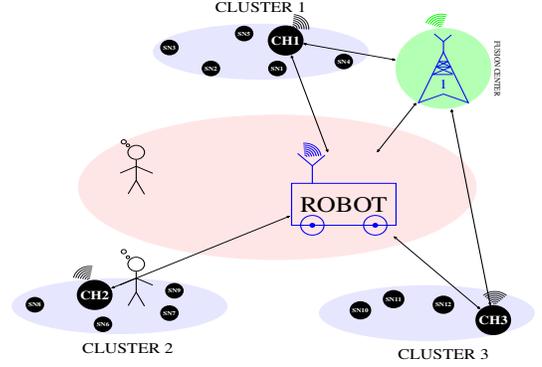}}}\vspace{-0.25cm}
    \caption[Schematic communication architecture among peripheral CHs]{\label{fig:Distributed_architectudsadasdasdsare}
  \small{ Schematic communication architecture among peripheral CHs, MR, and FC. The $i$th CH generates a test statistic ($T_i$) by combining the observations received from the SNs within the cluster. The CH can communicate with the FC directly or via the MR.}}
\end{figure} 
\section{Proposed Solution}
\label{Proposed Solution}
To extend the operational lifetime of the WSN and to deal with the imbalanced load among the CHs, we propose a $double-link$ mobility diversity algorithm and derive a MR path planner that we describe next in Section \ref{Multiple Link Mobility Diversity Algorithm}.
We assume that the FC, at time instant $t=k\tau$, has full knowledge of the channel gains ($h(\mathbf{q}_0,\mathbf{q}_j), \forall j=1,2, \ldots, N$) from CHs to FC. Based on this information, the FC determines the $L$ CHs with the lowest (CH-to-FC) channel gain and forward the corresponding CHs' identities to the MR. Then, the MR will act as a decode and forward relay that will establish communication links only with these $L$ (FC selected) CHs. Nevertheless, due to small-scale fading, these communication links may significantly reduce the communication quality and eventually a larger amount of CH transmit power may be required to satisfy the reference power at the receiver (\ref{eq:2.3}). Since the CH nodes are battery operated, optimizing their transmit power is of a particular importance in extending the WSN operational lifetime.

We note that after a duration $\tau$, the wireless communication links may change due to the temporal dependence of the small-scale fading term. As a result, the FC is required to continuously estimate the set of $L$ CHs with the lowest CHs-to-FC channel gains and forward the corresponding CHs’ identities to the MR. 

Before introducing the proposed algorithm, we next briefly describe few existing fading compensation techniques.
\subsection{Related Compensating Techniques}
\label{Related Algorithms}

Few existing techniques deal with the fading channel compensation in practice. A widely used technique is the multi-antenna diversity. In the context of WSNs, clearly this technique requires multiple antennae transceivers mounted on each of the CH node. This not only increases the cost of the node but also in many practical scenarios, may not be even feasible due to both the node size limitations and transmit power constraint. Hence, single antenna transceivers are desired in practice.

Now, since the small-scale fading term is time varying, temporal diversity technique could be used. But because this work consider the scenarios where the coherence time is significantly greater than the symbol duration, the temporal diversity is not suitable as it would introduce a large delay. In this case, the MDAs \cite{15} are suitable techniques to compensate the small-scale fading in WSNs.
\subsection{Multiple Link Mobility Diversity Algorithm}
\label{Multiple Link Mobility Diversity Algorithm}

MDAs are a new type of diversity technique that exploit the spatial variations of the small-scale fading and the mobility of the MRs. Their operation is divided in two phases\cite{15,16}: (i) exploration phase; and (ii) selection phase. During the exploration phase, the MR explores a series of $K$ stopping points located in its vicinity (from where it estimates the channel gain) that are optimized based on a path planner. After the exploration phase, the MR uses a selection rule to decide on the optimum position for establishing a communication link. 

The existing MDAs (e.g., \cite{15}) are only applicable to the compensation of a single small-scale fading channel. Here, we require a simultaneously small-scale fading compensation technique of $L+1$ communication links (i.e., the $L$ MR-to-CHs as well as the MR-to-FC links). Therefore, in this paper we extend our previous work in \cite{15,16} and develop a $multiple-link$ MDA. To the best of our knowledge, this is the first time that a $multiple-link$ MDA is proposed. Next, we develop the path planner that determines the location of the stopping points during the exploration phase and the MDA selection rule used during the selection phase. 
\subsubsection{MDA development}
\label{Multiple Link path Planner}

In the context of MDAs (as previously stated), the set of distances among all stopping points are small (see \cite{15}). As a result, the shadowing term of the $j$th CH-to-MR link at time $t$ is assumed to be constant, i.e., $s(\mathbf{p}(t),\mathbf{q}_j)\approx s_{j}, \forall j=1,2, \ldots, N$. Also, due to the fact the that distances between stopping points in MDAs are small then the distance travelled by the MR, while executing a n MDA, is on the order of a few wavelengths $\lambda$ and so it is much smaller than MR-to-FC and MR-to-CH distances. Therefore $\|\mathbf{p}(t)-\mathbf{q}_j\|_2\approx d_j$ for all $t$. 

In this paper, we develop a path planner with memory that uses channel gain measurements both at the current and previous MR's positions in order to estimate the MR's next position. This path planner requires small-scale fading predictors such as the one used in \cite{15}. Here, for simplicity, we choose the first order predictor\footnote{This predictor considers only the measurements of the channel at the current MR's position to predict the small-scale fading term at the next position.}. However, using the results presented here, the development of MDA with higher memory order predictors can be easily established. 

The small-scale fading predictor at time instant $t_{n+1}$ given the estimate ${\hat h}(\mathbf{p}(t_{n}),\mathbf{q}_j)$ is \cite{15}:
\begin{eqnarray}
\label{eq:3.1}
\tilde{h}(\mathbf{p}(t_{n+1}),\mathbf{q}_j)&=&\rho(\mathbf{p}(t_{n+1}),\mathbf{p}(t_{n}))\hat{h}(\mathbf{p}(t_n),\mathbf{q}_j)\\
&+&\left(\sqrt{1-\rho^2(\mathbf{p}(t_{n+1}),\mathbf{p}(t_{n}))}\right)u_{j,n}\nonumber
\end{eqnarray}
where $t_{n+1}-t_n\ll \tau$; $\hat{h}(\mathbf{p}(t_{n}),\mathbf{q}_j)$ is the estimate of ${h}(\mathbf{p}(t_{n}),\mathbf{q}_j)$, $\rho(\cdot)$ as in (\ref{eq:2.2}), and $u_{j,n}$ is a set of Normal independent and identically distributed random variables for $0\leq j\leq L$, $1\leq n\leq K$. 

To develop the path planner for the $multiple-link$ MDA, the MR position at time instant $t_{n+1}$ (i.e., $\mathbf{p}(t_{n+1})$), is chosen such that the minimum channel gain is maximized over $L+1$ links. So, our optimisation problem is:
\begin{equation}
\label{eq:3.88}
\begin{array}{l}
\displaystyle\mathrm{maximize}_{\ell_n\in[\ell_d,\ell_u]}\ \ G_1(\mathbf{p}(t_{n+1}))\\
\mathrm{s.t.}\\
\mathbf{p}(t_{n+1})=\mathbf{p}(t_{n})+\ell_n[\cos(\phi_n)\ \ \sin(\phi_n)]^T
\end{array}
\end{equation}
where 
\begin{equation}
\label{eq:3.2}
 G_1(\mathbf{p}(t_{n+1}))=\mathbb{E}\left[
\min_{j=0,1,\cdots,L}\left\{\frac{s_j\left|\tilde{h}(\mathbf{p}(t_{n+1}),\mathbf{q}_j)\right|} {d_j^{\alpha/2}}  \right\} 
\right]
\end{equation}
and $\ell_n$ is the distance travelled by the MR between the current and the next position, $\phi_n$ represents the MR movement direction, and finally $\mathbb{E}\left[\cdot\right]$ denotes the expected value with respect to the random variables set $\{u_{j,n}\}_{\forall j}$.  We would like to make it clear that $\ell_d$ is a design parameter and that $0<\ell_d<\ell_u$. Here, $\ell_u$ is the smallest distance $\ell$ such that $J_0(2\pi\ell/\lambda)=0$ (i.e., the smallest distance such that the small-scale fading terms in (\ref{eq:3.1}) are independent). Defining $\ell_n$ as above yields a normalized correlation factor $\rho(\mathbf{p}(t_{n+1}),\mathbf{p}(t_{n}))$ defined over the interval $[0,1)$.

We remind the reader that in (\ref{eq:3.2}), $\mathbf{q}_0$, denotes  the FC location. Since the predictor (\ref{eq:3.1}) is a complex Gaussian random variable then it can be easily shown that:
\begin{equation}
\label{eq:3.3}
G_1(\mathbf{p}(t_{n+1}))=\int_{0}^\infty\displaystyle\Pi_{j=0}^{L}Q_1\left(\frac{\nu_j}{\sigma_j},\frac{x}{\sigma_j}\right)\mathrm{d}x
\end{equation}
where $Q_1(\cdot,\cdot)$ is the Marcum Q function with
\begin{equation}
\label{eq:3.4}
\sigma_j=\frac{s_j\sqrt{1-\rho^2(\mathbf{p}(t_{n+1}),\mathbf{p}(t_{n}))}} {d_j^{\alpha/2}}
\end{equation}
\begin{equation}
\label{eq:3.5}
\nu_j=\frac{s_j\rho(\mathbf{p}(t_{n+1}),\mathbf{p}(t_{n}))|\hat{h}(\mathbf{p}(t_n),\mathbf{q}_j)|} {d_j^{\alpha/2}}.
\end{equation}
Solving the optimisation problem (\ref{eq:3.88}) is computationally expensive in general\footnote{The integral in (\ref{eq:3.3}) needs to be calculated numerically since it can not be evaluated analytically. This not only incur delays into the MDA algorithm but also require more processing power from the MR.}. Therefore we develop an alternative optimization problem which is similar but much simpler to solve. To do this, we first note that each multiplicative term in (\ref{eq:3.3}) is a monotonically decreasing function that  tends to zero. Then, there exists a value $X_0$ such that:
\begin{equation}
\label{eq:3.6}
\int_{0}^\infty\displaystyle\Pi_{j=0}^{L}Q_1\left(\frac{\nu_j}{\sigma_j},\frac{x}{\sigma_j}\right)\mathrm{d}x\approx\int_{0}^{X_0}\displaystyle\Pi_{j=0}^{L}Q_1\left(\frac{\nu_j}{\sigma_j},\frac{x}{\sigma_j}\right)\mathrm{d}x
\end{equation}
Using Chebyshev's inequality:
\begin{eqnarray}
\label{eq:3.7}
\displaystyle\frac{\int_{0}^{X_0}\Pi_{j=0}^{L}Q_1\left(\frac{\nu_j}{\sigma_j},\frac{x}{\sigma_j}\right)\mathrm{d}x}{X_0} &\geq & \frac{\displaystyle\Pi_{j=0}^{L} \int_{0}^{X_0} Q_1\left(\frac{\nu_j}{\sigma_j},\frac{x}{\sigma_j}\right)\mathrm{d}x}{X_0^L}\nonumber\\
&=&\frac{1}{X_0^L}G_2(\mathbf{p}(t_{n+1}))
\end{eqnarray}
with:
\begin{equation}
\label{eq:3.7b}
G_2(\mathbf{p}(t_{n+1}))\triangleq\displaystyle\Pi_{j=0}^{L} \left\{\sigma_j\sqrt{\frac{\pi}{2}}L_{1/2} \left(\frac{-\nu_j^2}{2\sigma_j^2}\right) \right\}
\end{equation}\\
where $L_{1/2}(\cdot)$ is Laguerre's polynomial of degree $1/2$. 

Then we obtain the alternative optimization problem by replacing in (\ref{eq:3.88}) the optimization target $G_1(\mathbf{p}(t_{n+1}))$ by its lower bound $G_2(\mathbf{p}(t_{n+1}))$:
\begin{equation}
\label{eq:3.8}
\begin{array}{l}
\displaystyle\mathrm{maximize}_{\ell_n\in[\ell_d,\ell_u]}\ \ G_2(\mathbf{p}(t_{n+1}))\\
\mathrm{s.t.}\\
\mathbf{p}(t_{n+1})=\mathbf{p}(t_{n})+\ell_n[\cos(\phi_n)\ \ \sin(\phi_n)]^T
\end{array}
\end{equation}
where $G_2(\mathbf{p}(t_{n+1}))$ is defined in (\ref{eq:3.7b}), $\ell_n$ is defined over the interval $[\ell_d,\ell_u]$ and determines the correlation between the small-scale fading terms (see (\ref{eq:2.2})).

From the extensive numerical results, we have observed that the optimisation problem (\ref{eq:3.8}) yields an optimum value $\ell^o_n$ either equal to $\ell_d$ or $\ell_u$ with a very high probability. Hence, to further simplify the optimization process and to reduce the MR processing burden, (\ref{eq:3.8}) is solved only for $\ell_n\in\{\ell_d,\ell_u\}$. It is worth noting that the optimisation is performed at time instant $t_{n}$ by making use of the observed communication channel measurements at MR position $(\mathbf{p}(t_{n}))$.

Clearly, solving the optimisation problem (\ref{eq:3.8}) will yield a set of stopping points with good wireless channel properties.  Now, the final step is to decide how to determine, among those stopping points, the optimum MR position such that the overall WSN performance is improved. In this paper, to achieve this the MR will select this optimum stopping point as the one that maximizes the minimum channel gain (i.e., incorporating all the MR-to-CH and the MR-to-FC links).

\section{Simulations}
\begin{figure}[htp!]              
\centerline{{\includegraphics[clip, trim ={4mm 40mm 1mm 50mm},width=75mm ,height=57mm]{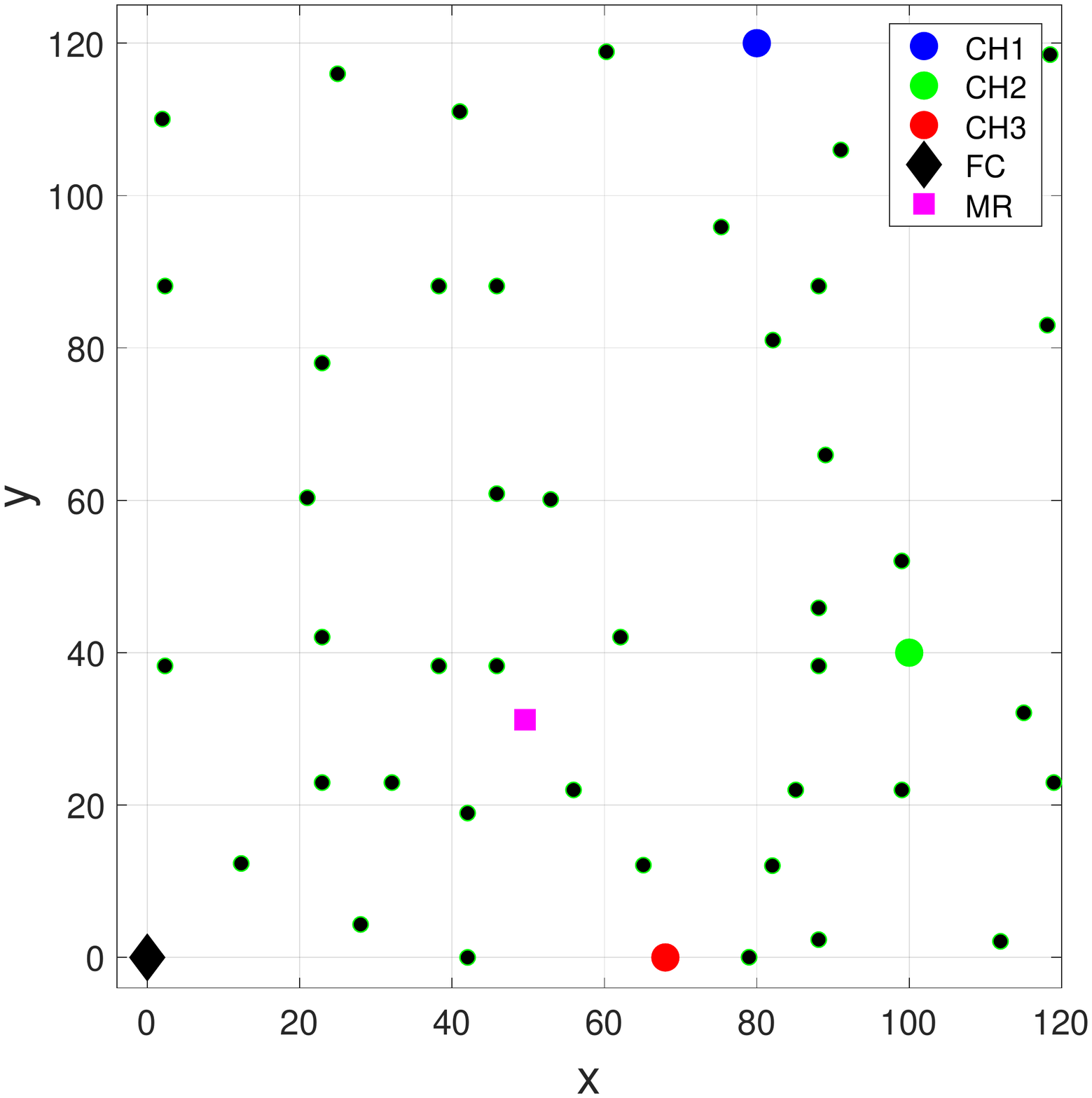}}}\vspace{-0.25cm}
    \caption[Global probability of detection for different P_t values]{
   \small{Spatial configuration of the WSN where the SNs are represented with green.}}
  \label{Figure555}
\end{figure}

We evaluate numerically the performance of our proposed $multiple-link$ MDA. We simulate a WSN deployed in a $120\times120$ ROI and $M$ SNs divided into $N=3$ clusters with arbitrary SN geometry, where the distances between the MR and CHs are assumed to be known. The spatial configuration is shown in Fig. \ref{Figure555}. 
\begin{figure}[htp!]              
\centerline{{\includegraphics[clip, trim ={-3mm 45mm -5mm 50mm},width=90mm ,height=90mm]{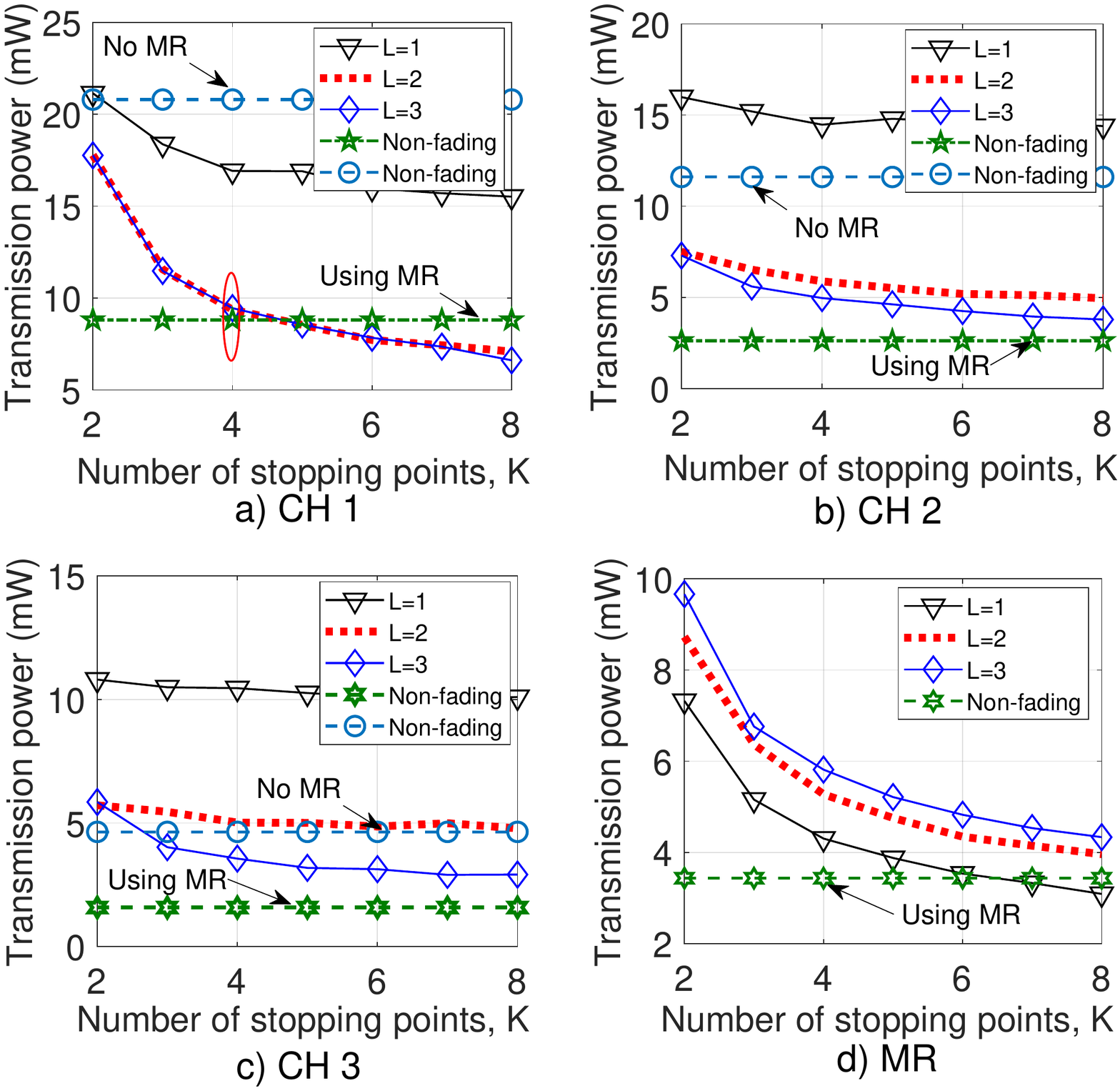}}}\vspace{-0.25cm}
    \caption[Global probability of detection for different P_t values]{
  \small{Average CHs transmission power in (3)  versus the number of stopping points ($K$), parametrized on the number of CHs that use the MR as a relay ($L$) with $P_{ref}=1$ $\mu$W, and $\alpha=2$.}} 
  \label{Figure4}
\end{figure}
We let the pathloss coefficient $\alpha=2$, the variance of $s_j(\mathbf{p},\mathbf{q}_j)$ in (\ref{eq:2.3}) is taken such that $\mathrm{Var}\left\{10\log_{10}(s_j(\cdot))\right\}=1$dB, $\forall j$ and the reference power $P_{ref}=1\mu$W. Finally, we note that both $G_1(\mathbf{p}(t_{n+1}))$ and $G_2(\mathbf{p}(t_{n+1}))$ are independent of the MR movement direction and we take $\phi_n=\phi$, $\forall n$ in (\ref{eq:3.8}).

In Fig. \ref{Figure4}, we show the mean MR and CHs transmit power (after executing the MDA) for different values of $L$ compared to the $non-fading$ case\footnote{In the $non-fading$ case, the pathloss only communication channel is considered and the MR is located at its initial point as in Fig. \ref{Figure555}.}. Clearly, as expected, as the number $K$ of stopping points increases, the MR's probability of finding a stopping point with a large channel gain also increases. As a result, the CHs transmit power decreases. 

Regarding the number of links $L$ considered by the MR during the MDA execution we observe that the MR transmit power increases when  $L$ increases (see Fig. \ref{Figure4} sub-figure d). This is expected as an increase in $L$ will cause the MR  to consider a larger number of communication links. Hence, decreasing the degree of freedom to obtaining simultaneously large channel gains for both itself and the $L$ CHs. 

Interesting, in Fig. \ref{Figure4} (i.e., sub-figure a)), we can observe that when the MDA is used, the  CH1 transmit power is lower than the $non-fading$ case when the number of stopping points $K$ is larger than 4 and for $L=2$ and $L=3$. This is due to the fact that the MDA takes advantage of the fading to improve the channel gain.

\begin{figure}[h!]              
\centerline{{\includegraphics[clip, trim ={-8mm 50mm 10mm 59mm},width=75mm ,height=54mm]{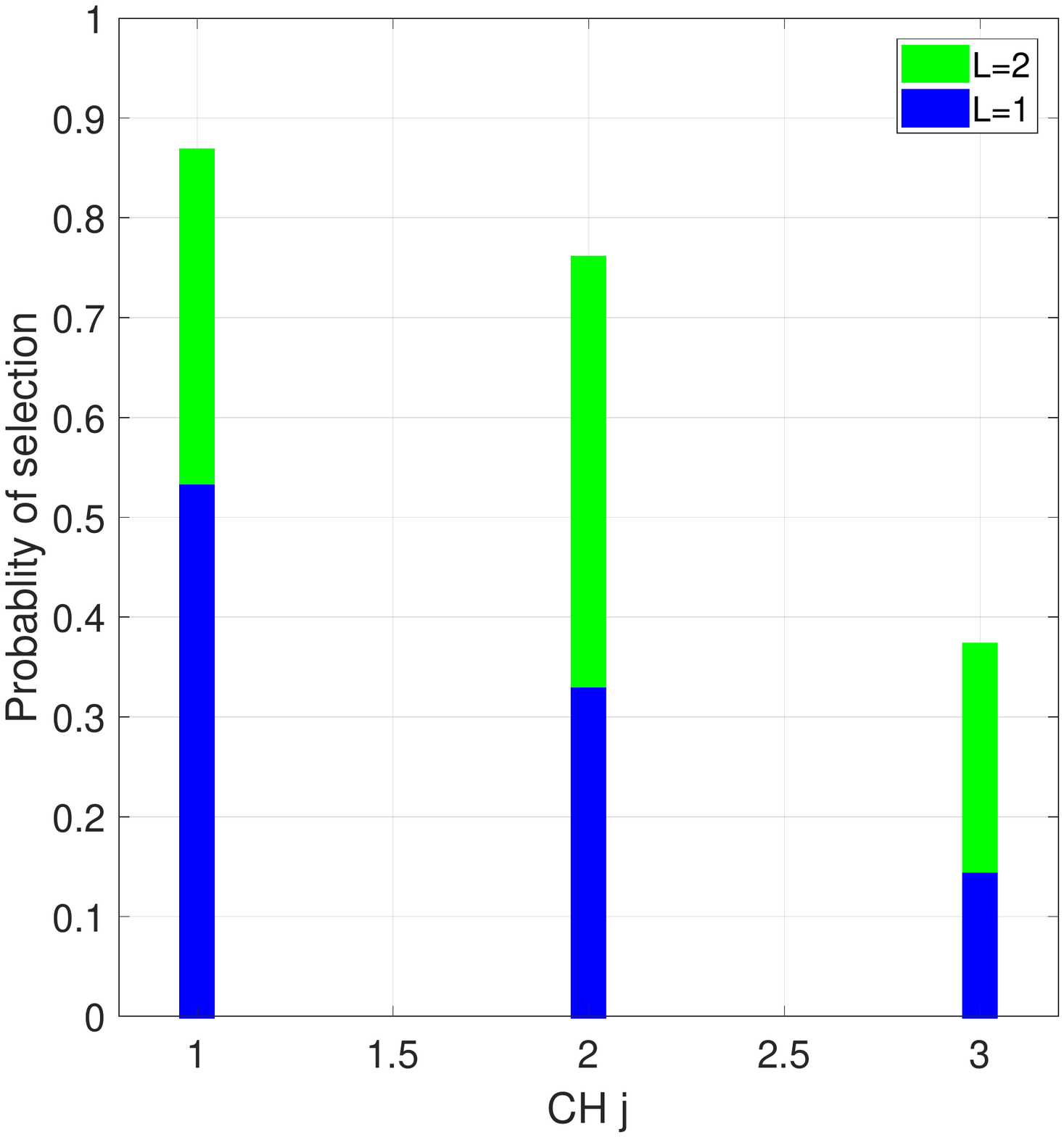}}}\vspace{-0.25cm}
    \caption[Global probability of detection for different P_t values]{
  \small{CH's selection probability to use the MR as a relay versus the CH ($j$), parametrized on the number of CHs that use the MR as a relay ($L$) with $P_{ref}=1$ $\mu$W, and $\alpha=2$.}}
  \label{Figure5}
\end{figure}

Now, to further validate our results, in Fig. \ref{Figure5} we observe the CH selection probability (by the MR) for $L=1$ and $L=2$ (when $L=3$ all CHs are selected). Clearly, there is a correlation between the selected CHs (i.e., the $L$ CHs with the lowest channel gain) and the corresponding selection probability. For example, CH1 (experiencing the worst communication channel) has a higher selection probability than CH2 and CH3. Furthermore, we can observe an increase in selection probability for all CH as $L$ increases. This is as expected since an increase in $L$ also increases the opportunity of a particular CH to use the MR as a relay in improving its communication link. Therefore as $L$ increases each CH will have a higher opportunity to benefit from the MDA. This is why the transmit power for the CHs decreases as $L$ increases (see Fig. \ref{Figure4}). 

\section{Conclusions}
\label{conclusions}
In this paper, we propose an efficient $multiple-link$ MDA to balance the CHs energy and extend their operational lifetime in random clustered WSNs. We have shown how by using an MR as a relay with the proposed MDA, the CH's mean transmit power can be significantly reduced. Finally, we have also shown that the proposed MDA results in a lower CH's transmit power compared  to the  non-fading communication channel case. Future work will investigate the analysis of the problem for fully distributed solution (i.e., where there is no FC).

\end{document}